# Gate-Controlled Magnetoresistance of a Paramagnetic Insulator|Platinum Interface

L. Liang[1], J. Shan[1], Q. H. Chen[1], J. M. Lu[1,†], G. R. Blake[1], T. T. M. Palstra[1,‡],

G. E. W. Bauer[1,2,*], B. J. van Wees[1] and J. T. Ye[1,*]

[1] *Zernike Institute for Advanced Materials, University of Groningen, Nijenborgh 4, 9747 AG, Groningen, The Netherlands*

[2] *IMR & WPI-AIMR & CSRN, Tohoku University, Sendai 980-8577, Japan*

[†] *Present address: State key laboratory for mesoscopic physics, Peking University, 100871 Beijing, People's Republic of China*

[‡] *Present address: College van Bestuur, Universiteit Twente, 7500 AE, Enschede, The Netherlands*

We report an electric field-induced in-plane magnetoresistance of an atomically flat paramagnetic insulator|platinum (Pt) interface at low temperatures with an ionic liquid gate. Transport experiments as a function of applied magnetic field strength and direction obey the spin Hall magnetoresistance phenomenology with perpendicular magnetic anisotropy. Our results establish the utility of ionic gating as an alternative method to control spintronic devices without using ferromagnets.

\* Corresponding authors: j.ye@rug.nl, g.e.w.bauer@imr.tohoku.ac.jp

Magnetoresistance (MR), the change of the electrical resistance by external magnetic fields, is the key functionality of data storage [1,2], sensors [3] and logic devices [4]. The evolution of information technology relies on the discovery of new types of MR. For example, the giant magnetoresistance [1,2] and tunnel magnetoresistance [5,6] in magnetic multilayers have been breakthroughs in the field of spintronics that triggered technological revolutions. Conventional magnetoresistive devices contain ferromagnetic elements with stray fields that cause undesirable cross-talk energy loss. Paramagnets that lack spontaneous magnetization play only passive roles, *e.g.* as spacer layers [7]. The magnetization of ferromagnets cannot be simply switched off; even the physically important and its technologically desirable electric control [8–10] is difficult due to the intrinsically large carrier density and consequently short Thomas-Fermi screening length of metallic magnets. These drawbacks can be overcome by ionic gating which can generate very large electric fields by applying only a few volts [11–13].

Platinum (Pt) is an essential material for spintronics. It is widely employed as spin injector and detector due to its strong spin-orbit interaction (SOI) and hence large spin Hall angle [14]. According to the Stoner criterion, the large density of state at the Fermi energy puts Pt very close to the ferromagnetic (FM) phase transition. Recently, ferromagnetism was induced in Pt by electrostatic gating using paramagnetic ionic liquid (PIL) [15], a special type of ionic liquid containing paramagnetic ions. The gate-induced carriers are confined on an atomic length scale to the Pt surface. The PIL on top of Pt forms an atomically flat interface to Pt and becomes an electrical insulator below its melting point ($T_m$). In contrast to conventional magnetic thin film multilayers, the physical properties of the present interface can be tuned by varying the voltage of the PIL gate in its liquid phase. Here, we report the observation of a novel gate-controllable MR in the PIL|Pt system for an in-plane magnetic field **B**. We find that the gating induces a resistance that depends on the direction of **B**. The symmetry is distinctively different from the conventional anisotropic magnetoresistance (AMR) of ferromagnets or the spin Hall magnetoresistance for Pt|magnetic insulator bilayers with in-plane magnetizations [14,16,17]. On the other hand, the observations can be well explained by the spin Hall magnetoresistance when the conduction electrons in the bulk Pt interacts with the interface with perpendicular magnetization. The results illustrate the unique tuning option provided by our system that adds functionalities to spintronic devices such as easy reprogrammability.

The transport properties are measured in a PIL-gated transistor shown schematically in Fig. 1a [18]. Our device consists of a Pt Hall bar ($t$ = 12 nm) covered by a PIL gate. The PIL used in our experiment is butylmethylimidazolium tetrachloroferrate (BMIM[FeCl$_4$]) (Fig. 1b inset). The *d*-shell of $Fe^{3+}$ in the magnetic anions is half-filled with spin quantum number $S$ = 5/2 (high spin state) [19]. Magnetic susceptibility ($\chi_m$) measurements of BMIM[FeCl$_4$] follows Curie's law indicating BMIM[FeCl$_4$] is paramagnetic (PM) (Fig. 1b) with a large effective magnetic moment $\mu$ = 5.77 $\mu_B$, where $\mu_B$ is the Bohr magneton. Assuming orbital quenching the magnetization agrees well with the theoretical value for a half filled 3$d$ atomic shell of 5.92 $\mu_B$ calculated from $g\sqrt{S(S+1)}$, where $g$ = 2 is the Landé factor. By varying the gate voltage $V_G$, cations or anions are driven to the Pt surface, which collects or depletes electrons in the top-most layer, respectively.

The $V_G$ dependence of the longitudinal resistance $R_L$ shows reversible control with negligible leakage current $I_G$ (Fig. 1c), which indicates a change of the electronic surface state of Pt. At positive $V_G$, $R_L$ decreases with the increase of Fermi level in the band structure of Pt [15] and saturates after $V_G$ > 2 V. We fix this low resistance state at $V_G$ = 2.2 V ("ON" state) by rapidly cooling the device below the $T_m$ of the PIL. Angular-dependent magnetoresistance (ADMR) measurements at 5 K show a clear modulation of $R_L$ (red curves in Fig. 1d,e), indicating a dramatic change of the magnetic properties of the Pt channel after switching to the "ON" state. This effect disappears after subsequent release of $V_G$ at 220 K and cool-down to the same temperature ("OFF" state) (blue curves in Fig. 1d,e). This direct correlation of $\Delta R_L$ as a function of $\phi$ (Fig. 1c) with $V_G$ proves that the observed effect is induced by the PIL gating. We also find that gating by a non-paramagnetic liquid (*e.g.* N,N-diethyl-N-methyl-N-(2-methoxyethyl)ammonium bis(trifluoromethanesulfonyl) imide: DEME-TFSI) does not lead to such a state [15].

Fig. 2b shows ADMR measurements of the longitudinal resistivity $\rho_L$ at 5 K for the "ON" state under **B** field strengths from 0.5 T to 6 T. All measurements display harmonic modulations with a period of $\pi$, where the maximum and minimum values for **B** || **I** (current) and **B** ⊥ **I** are denoted as $\rho_\parallel$ and $\rho_\perp$, respectively. A similar angle dependence is observed for the transverse resistivity $\rho_T$, in which the maxima and minima are shifted by 45° (Fig. 2c). This dependence mimics the AMR and planar Hall effect (PHE) of ferromagnets, in which $\rho_L$ and $\rho_T$ are governed by the angle $\phi$ between **I** and magnetization **M** as

$$\rho_L(\phi) = \rho_\perp + \Delta\rho_L \cos^2 \phi \quad (1)$$

$$\rho_T(\phi) = \Delta\rho_T \sin\phi \cos\phi . \quad (2)$$

Despite the similarity in shape, our field-dependent amplitudes $\Delta\rho_L = \rho_\parallel - \rho_\perp$ and $\Delta\rho_T = \rho(\phi=45°) - \rho(\phi=135°)$ is significantly different from the AMR. The field-dependent modulations in the present system show the same values of $\rho_L$ at $\phi$ = 0°, indicated by the dashed lines (Fig. 2b), whereas $\rho_L$ would be constant at $\phi$ = 45° for the latter case. Moreover, if the effect is caused by AMR, $\Delta\rho_L$ ($\Delta\rho_T$) and M = |**M**| saturate for fields exceeding the coercivity $\mu_0 H_c$ and remain finite even at $B$ = 0 T due to magnetic remanence. We, however, find $\Delta\rho_L$ ($\Delta\rho_T$) to vanish without **B** and increase with field strengths without saturating even at 6 T (Fig. 2d). This dependence resembles the magnetization curve of the PIL at 5 K, which can be well-described by the Langevin function of paramagnetism

$$L(x) = \coth x - 1/x, \tag{3}$$

where $L(x) = \frac{M}{M_s}$, $x = \frac{\mu B}{k_B T}$, $M_s$ is the saturation magnetization, $\mu$ the magnetic moment, and $k_B$ the Boltzmann constant.

We further characterize $\Delta\rho_L$ ($\Delta\rho_T$) by the field-dependent magnetoresistance (FDMR) at the same temperature ($T = 5$ K) and at various angles $\phi$. At $\phi = 90°$, when **B** is in the sample $xy$-plane and perpendicular to the current direction **I**, we observe a negative MR that does not saturate at the maximum **B** field of 6 T (Fig. 2e, f). At $\phi = 0°$, when **B** is in-plane and along **I**, however, the MR vanishes for all **B** fields. Although in the ADMR measurement, the $\rho_L$ profile at a fixed **B** strength shows an AMR-like modulation ~**M·I**, the observed anisotropic FDMR firmly excludes the AMR.

With increasing temperature, the magnetic susceptibility of the PIL decreases significantly (Fig. 1b) and the induced magnetization of the Pt surface becomes weaker. Both effects are expected to affect the interface MR. We therefore carry out FDMR measurements at various temperatures under both in-plane and out-of-plane **B** fields.

Figs. 3a,b show the evolution of the in-plane MR with increasing temperature. All data were collected at $\phi = 45°$, at which angle both $\rho_L$ and $\rho_T$ depend on |**B**|. By symmetry, $\delta\rho_L(45°) = \rho_L(B) - \rho_L(0)$ should be $\sqrt{2}/2$ times of $\delta\rho_L(90°)$ that is equal to $\Delta\rho_L = \rho_\parallel - \rho_\perp$ in the ADMR measurement; moreover $\delta\rho_T(45°) = \rho_T(B) - \rho_T(0)$ should equal $\Delta\rho_T$ from the ADMR. For comparison, we also measured the longitudinal (magnetoconductivity $\sigma_P$) and transverse signals (Hall resistivity $\rho_H$) as a function of perpendicular **B** in the out-of-plane geometry at different temperatures, as shown in Fig. 3c, d. Similarly, the changes of the interface $\sigma_P$ are defined as $\delta\sigma_P = \sigma_P(B) - \sigma_P(0)$ after subtracting the parallel bulk contribution; whereas $\delta\rho_H$ is extracted by extrapolating the linear Hall response from high to zero magnetic field (red lines in Fig. 3d). A crossover of $\delta\rho_L$, $\delta\sigma_P$ and $\delta\rho_H$ from negative values at low temperature to positive ones at high temperature causes a sign change at roughly 40 K (Fig. 3a,c,d). In contrast, $\delta\rho_T > 0$ up to the highest measured temperatures (Fig. 3b).

In a conductor with spontaneous **M**, Ohm's relation between the electric field **E** and the electric current **I** reads [20]

$$\mathbf{E} = \rho_\perp \mathbf{I} + (\rho_\parallel - \rho_\perp)(\mathbf{M} \cdot \mathbf{I})\mathbf{M} + \rho_H \mathbf{M} \times \mathbf{I}. \tag{4}$$

When **B** is perpendicular to the sample $xy$-plane (hence the direction of **I**), we detect an anomalous transverse voltage (also at 5 K) (Fig. 3d), *i.e.* last term in Eq. (4), where $\rho_H$ is the Hall resistivity. The observation is consistent with our recent report of a PIL gating-induced FM state in Pt with perpendicular magnetic anisotropy (PMA) [15]. The effect of an in-plane **B** on the intrinsic perpendicular **M** can be described by the Stoner-Wohlfarth model of coherent magnetization rotation [21].

The **B** dependence of the **M** of the PIL show no hysteresis even at the lowest temperatures (Fig. S1), indicating the absence of long-range FM ordering. Therefore, the **M** of the PIL increases with increasing magnetic field or decreasing temperature. Once being frozen, PIL becomes highly insulating and no electrical current can enter. In addition, the strong Thomas-Fermi screening limits gate-induced changes in the electronic state of the Pt to the top-most atomic layers [22,23]. Therefore, our paramagnetic gating-induced magnetoresistance (PMR) effect must originate from the Pt side of the PIL|Pt interface.

The large spin Hall angle of Pt [24] is known to convert a electrical current **I** efficiently into a transverse spin current with direction **I**$_s$ and polarization **σ** ∥ **I**×**I**$_s$. The magnetotransport in Pt contacts to conventional magnetic insulators is well-explained by the spin Hall magnetoresistance (SMR) model [16,17], which also appears to be consistent with many features of our experiments. Pristine Pt is a normal metal so an in-plane **B** field should not generate a significant MR (Fig. 1d, e and Fig. 4a). The spontaneous magnetization **M** at and perpendicular to the PIL|Pt interface when cooled down to low temperatures with $V_G$ is normal to the polarization of the spin motive force **σ** over the remainder of the Pt film. Without an external **B** field, **σ**⊥**M** and the generated spin current is efficiently absorbed as a spin transfer torque, leading to the high resistance state of $\rho_L$ (Fig. 4b). Increasing the in-plane **B** gradually pulls **M** into the plane. At $\phi = 0°$, **σ** is still normal to **M**, the absorption of the spin current remains constant, and $\rho_L$ remains at the high resistance state (Fig. 2e and Fig. 4c). At $\phi = 90°$, on the other hand, **M** is pulled into the plane with increasing **B** until **M** ∥ **σ**, at which the spin current generated by the SHE is mostly reflected, resulting in a decrease of $\rho_L$ by the inverse SHE (Fig. 2e and Fig. 4d). According to this SMR mechanism, the maximum amplitude of the longitudinal and transverse ADMRs should be the same $\Delta\rho_L = \Delta\rho_T$ when the sample geometry is factored in (Fig. 2a), in good agreement with low temperature data [25]. We adopt the highest SMR value $\Delta\rho/\rho = 0.027\%$, obtained with an external field of 6T at 5 K into the equation $\frac{\Delta\rho}{\rho} = \frac{\theta_{SH}^2}{t} \frac{4\lambda^2 G_r \tanh^2\frac{t}{2\lambda}}{\sigma+2\lambda G_r \coth\frac{t}{2\lambda}}$, where the thickness $t = 12$ nm, resistivity $\rho = 59.4$ μΩ cm, spin Hall angle $\theta_{SH} = 0.044$ and spin diffusion length of Pt $\lambda = 3.5$ nm [26], and find that $G_r = 2.88 \times 10^{14}$ S/m$^2$. This value is roughly in the same order of magnitude as that of the prototypical YIG/Pt systems.

The MR as a function of temperature is, on the other hand, governed by the competition between an MR that is diminished with the reduced spontaneous magnetization and an enhanced paramagnetism that generates magnetic order along an applied magnetic field. With increasing temperature, a positive MR effect evolves for both in-plane and out-of-plane **B** configurations in Fig. 3a and c, which possibly has a paramagnetic origin [27]. For the in-plane configuration and at an angle of $\phi=45°$, the FDMR (blue arrows or dots in Fig. 5) consists of the PMR effect on top of a positive shift of the background. At low temperatures, the strong PMA dominates and causes a negative MR (Fig. 5a). With the increase of temperature, the PMA weakens and the background that contributes to the observed signal increases. At ~40 K, the in-plane transport properties resemble the conventional AMR or SMR type of behavior with vanishing MR at $\phi = 45°$ (Fig. 5b). At even higher temperatures, the aforementioned positive MR eventually overwhelms the contribution of PMR, resulting in a positive FDMR signal (Fig. 5c). The transverse transport (Fig. 5d-f), on the other hand, remains almost unchanged at different temperatures, which is consistent with the following scenario. With increasing temperature, the PMA and **M** of the gate-induced FM layer weakens, leading to a gradual transition of the in-plane component of magnetization from a FM interface with PMA to one with enhanced PM susceptibility. Consequently, the PMR changes into the conventional in-plane SMR (Fig. 5f). When the in-plane **B** field forces the perpendicular magnetization into the plane at low temperature and that of polarizing the paramagnet at high temperatures are the same, the transverse magnetoresistance does not depend on temperature, as observed.

In conclusion, our study introduces a tunable spintronic system that employs a liquid paramagnetic insulator. At low temperatures, we observe a spin-dependent in-plane MR effect that can be explained by extending the SMR model to a PIL|Pt interface with PMA. The physics underlying the rich transport features as a function of temperature remains to be fully understood. The versatile gate-tunable magnetic phenomena lay the foundation for reprogrammable spintronic devices.

**Acknowledgments**

We would like to acknowledge J. Bass, J. Harkema, M. de Roosz and J.G. Holstein for technical assistances. This work is supported by ERC Ig-QPD grant, Ubbo Emmuis scholarship from University of Groningen, NanoLab NL, the Zernike Institute for Advanced Materials, and Grant-in-Aid for Scientific Research (Grant No. 26103006) of the Japan Society for the Promotion of Science (JSPS).

**Figures and Captions**

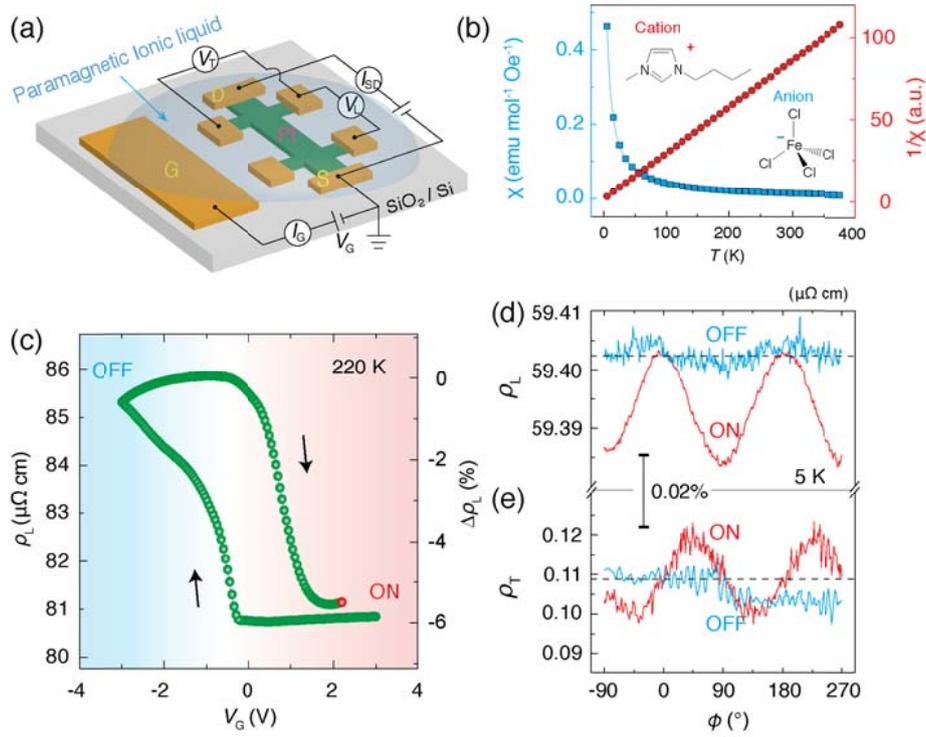

FIG. 1 (a) Schematic representation of our paramagnetic ionic liquid (PIL) gated platinum (Pt) thin film electric double-layer transistor. (b) Magnetic susceptibility $\chi$ of BMIM[FeCl$_4$] (green) measured under $B = 10$ mT. The linear behavior of $1/\chi$ (red) is indicative of Curie paramagnetism. Inset: Molecular structure of the paramagnetic ionic liquid 1-butyl-3-methylimidazolium tetrachloroferrate (BMIM[FeCl$_4$]) used as gating medium. (c) The gate dependence of the longitudinal resistance $R_L$ of a PIL-gated Pt, measured at 220 K with a gate voltage $V_G$ sweep rate of 20 mVs$^{-1}$. (d)-(e) The ADMR results of longitudinal and transverse magnetoresistances measured at 5 K with and without $V_G$. The ON and OFF states correlate with the gate dependence of the sheet resistance (Fig. S2) and are also labelled in panel (c).

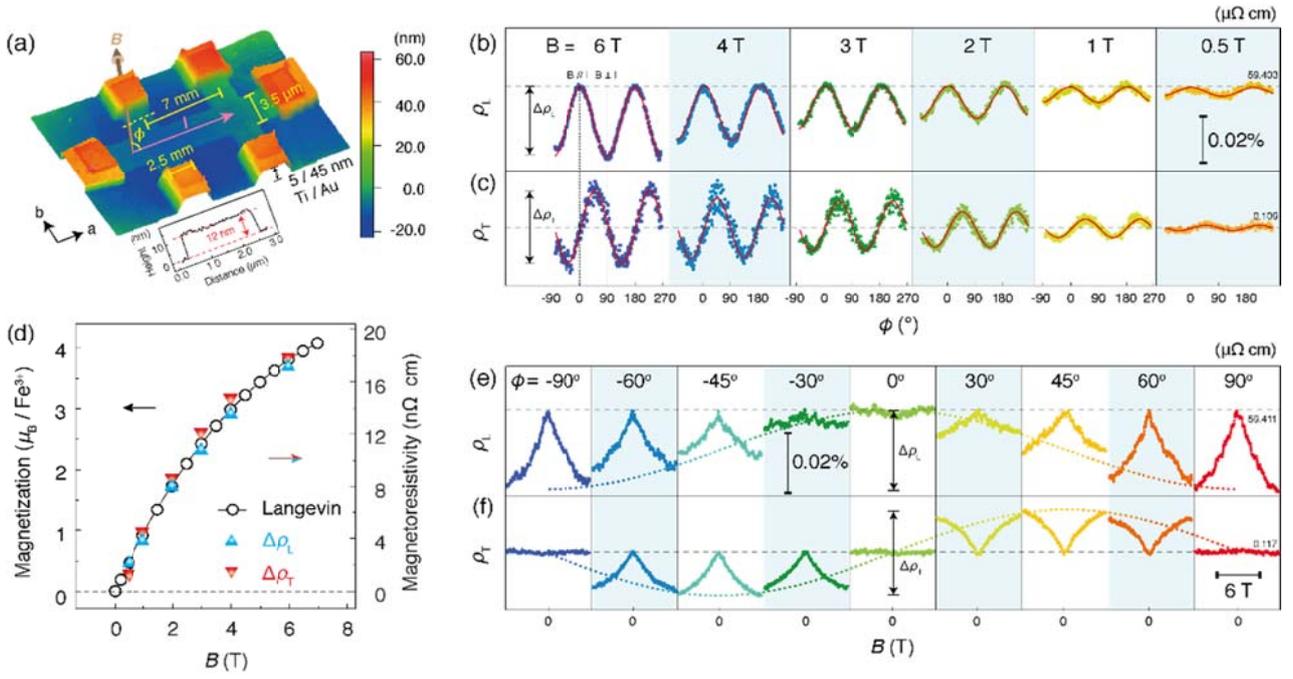

FIG. 2 (a) Atomic force microscope image of the device with bird-eye view. The height profile is illustrated with scale bar on the right. The thickness of the Pt film is 12 nm. The color bar indicates the height information. (b)-(c) In-plane magnetic field dependence of the longitudinal $\rho_L$ (b) and transverse $\rho_T$ (c) resistivities as a function of angle $\phi$ between magnetic field **B** and current direction **I**, measured at $T$ = 5 K. (d) Correlation between the magnetization of the PIL and in-plane longitudinal magnetoresistivities of the Pt thin film at $T$ = 5 K. Both $\rho_L$ (green) and $\rho_T$ (red) follow the magnetization curve of the PIL (in blue) measured directly by a Quantum Design SQUID. (e)-(f) Longitudinal (e) and transverse (f) magnetoresistivities for different in-plane magnetic field angles $\phi$ at 5 K. The black dashed lines indicate reference values at $B$ = 0 T. The colored dotted lines sketch the harmonic dependence of $\Delta\rho_L$ and $\Delta\rho_T$ on $\phi$ at $B$ = 6 T.

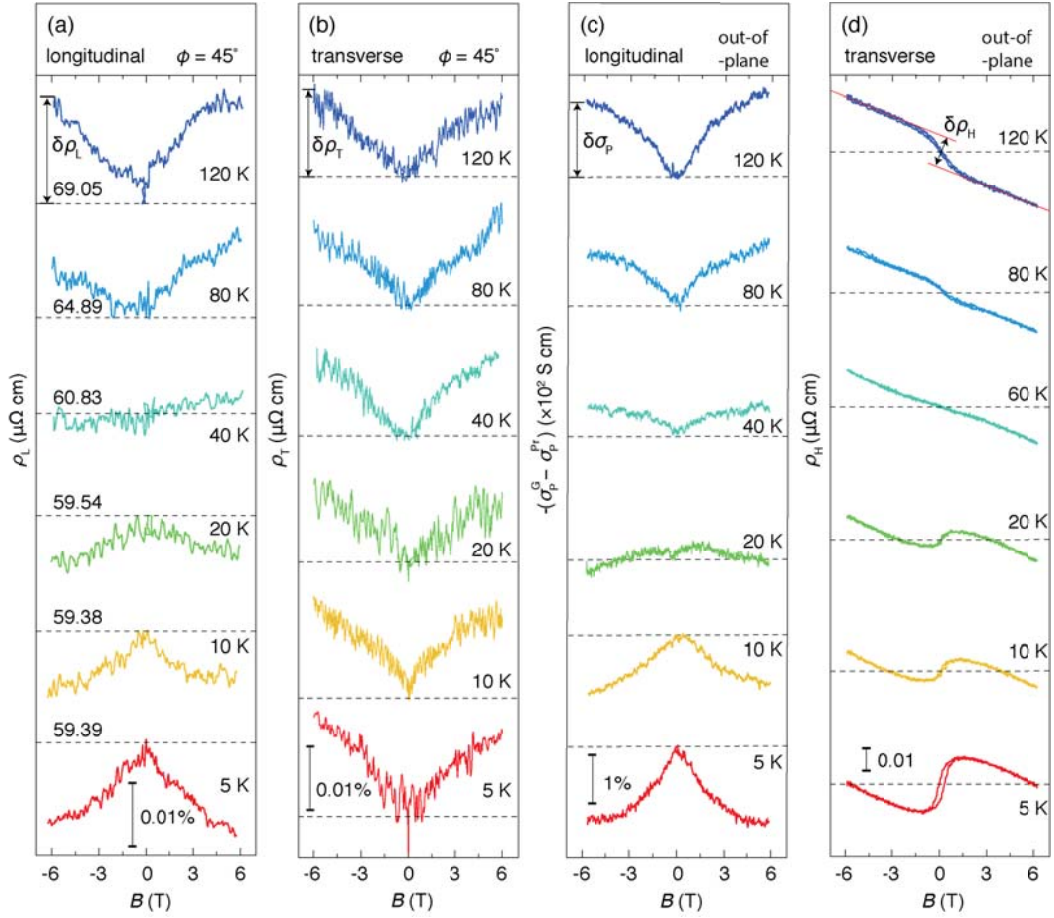

FIG. 3 Temperature-dependent (a) longitudinal and (b) transverse magnetoresistivities as a function of in-plane **B** field at an angle $\phi=45°$. (c) Longitudinal conductivities with out-of-plane **B** after subtracting the bulk contribution. (d) Anomalous Hall effect. The dashed lines represent the reference values for |**B**|=0; their values (if non-zero) are indicated just above the line. The temperatures are noted on the right of each panel.

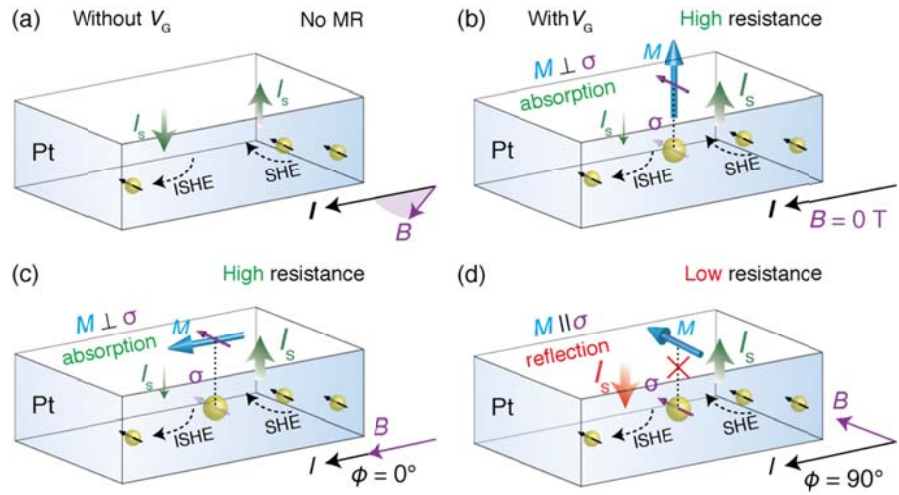

FIG. 4 Mechanism behind the spin-Hall magnetoresistance at a PIL|Pt interface for (a) without and (b-d) with gating. After gating by the PIL, the interface Pt becomes ferromagnetic with perpendicular magnetization. An applied **B** field forces the magnetization into the plane with high and low resistance states for **B** ∥ **I** and **B** ⊥ **I**, respectively.

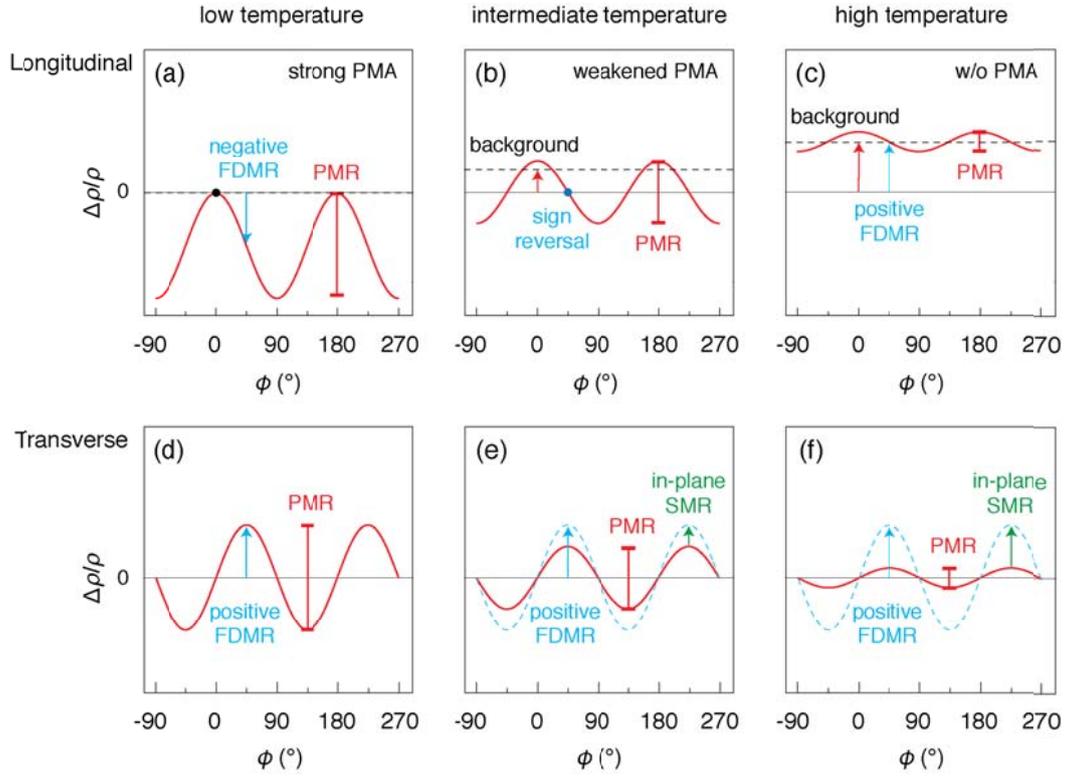

FIG. 5 Schematic of the temperature dependence of in-plane MR. (a)–(c) and (d)–(f) represent the longitudinal and transverse MR, in which $\Delta R/R = R(\phi,B)/R(\phi,B=0) - 1$. (a)–(d), (b)–(e) and (c)–(f) are at low, intermediate and high temperatures, respectively. The changes of the measured FDMR signals and the magnitudes of the PMR are illustrated by the blue arrows and red bars with respect to other effects.